\documentclass[usenatbib]{mn2e}

\usepackage{graphicx}
\usepackage{natbib}

\newcommand{\phoebe}{P{\sevensize HOEBE}}

\newcommand{\vizier}{V{\sevensize IZIE}R}

\begin{document}
\bibliographystyle{astron}

\title[Observations and Analysis of $\mu^{1}~{\mbox{Sco}}$]{New Observations and Analysis of the Bright Semi-Detached Eclipsing Binary \boldmath{$\mu^{1}~{\mbox{Sco}}$}}
\author[C.~van~Antwerpen and T.~Moon]{C.~van~Antwerpen$^1$ and T.~Moon$^2$ \\
        $^1$School of Chemistry, Physics and Earth Sciences, Flinders University, GPO~Box~2100, Adelaide SA 5001, Australia.\\\hspace*{2mm} e-mail: coenva@optusnet.com.au \\
        $^2$Sydney Institute for Astronomy, School of Physics, A28, The University of Sydney, NSW 2006, Australia.\\\hspace*{2mm} e-mail: T.Moon@physics.usyd.edu.au}

\maketitle
\begin{abstract}
Using new and published photometric observations of $\mu^{1}~{\mbox{Sco}}$~(HR~6247), spanning 70~years, a period of 1.4462700(5)~days was determined. It was found that the epoch of primary minimum suggested by Shobbrook at HJD~2449534.178 requires an adjustment to HJD~2449534.17700(9) to align all the available photometric datasets. Using the resulting combined-data light-curve, radial velocities derived from IUE data and the modelling software \phoebe{}, a new system solution for this binary was obtained. It appears that the secondary is close to, or just filling, its Roche-lobe.
\end{abstract}

\begin{keywords}
binaries: eclipsing
\end{keywords}

\section{Introduction}

$\mu^{1}~{\mbox{Sco}}$ (HR~6247; HD~151890; HIP~82514) was only the third spectroscopic binary to be discovered and is listed in the GCVS \citep{Samus04} as an eclipsing binary variable. Over the intervening years there have been a number of detailed measurements and several significant studies. It is now classified as a semi-detached~(sd) binary as one component is believed to fill its Roche lobe. \citet{CesterFedel77} considered $\mu^{1}~{\mbox{Sco}}$, to be unusual owing to:
\begin{enumerate}
\item their determination of an apparently high mass-ratio;
\item indications that the secondary component has a larger radius than the primary and overflows its Roche lobe;
\item both components appear to lie on the main sequence;
\item it being a member of the Scorpius-Centaurus cluster thus indicating an age of no more than $10^7$~years hence the possibility this system has just arrived on the main sequence.
\end{enumerate}

A comprehensive analysis of $\mu^{1}~{\mbox{Sco}}$ was undertaken to determine the best estimate of period from all available photometric data, establish a suitable epoch and calculate the key parameters defining the system.

The photometric data used included published measurements from:
\begin{enumerate}
\item \citet{RudnickElvey38},
\item \citet{vanGent39},
\item \citet{Stibbs48},
\item Tycho collected between~1990 and~1993 \citep{ESA97,OchsenbeinBauer00},
\item Hipparcos collected between~1990 and~1993 \citep{ESA97,OchsenbeinBauer00},
\item \citet{Shobbrook04},
\end{enumerate}
and new measurements taken in 2006, 2007 and 2008 by one of the authors~(Moon).

The photometric data was then combined with radial velocities obtained by \citet{SticklandSahade96} from IUE spectral data~\citep{IUE09} and the resulting dataset analysed using the software program \phoebe{} \citep{Prsa03,PrsaZwitter05c,PrsaGuinan08,Phoebe09} based on the established \citet{WilsonDevinney71} theoretical construct \citep{KallrathMilone99}.

\section{Published data}

The star $\mu^{1}~{\mbox{Sco}}$ was first reported to be a spectroscopic binary by \citet{Bailey96}. \citet{Maury20} undertook an extensive study of this star using 184~spectra collected over the period~1892 to~1918 and obtained a radial velocity curve of one component with respect to the other. From these data she determined a period of 1.44627~days and selected an epoch for the zero-point of the recession of the brighter component of the binary at HJD~2412374.434. The total light variation was between~0.3 and~0.4 of a magnitude and the combined mass of the system was estimated by her to have a minimum of 16.5~M$_{\sun}$.

\citet{RudnickElvey38} made 128~photoelectric measurements of $\mu^{1}~{\mbox{Sco}}$ using a photometer based on a Kunz $KH$~photocell as described by \citet{Stebbins31}. This cell had peak sensitivity near 460~nm and was typically used in conjunction with blue and yellow filters to define band-passes with effective wavelengths around 400~nm and 500~nm. Rudnick and Elvey do not, however, indicate the band-pass for their measurements or whether filters were used. Using their photometric measurements with the spectroscopic elements given by \citet{Maury20}, they determined the mass of the system to be 24.0~M$_{\sun}$, consistent with the minimum value of 16.5~M$_{\sun}$ determined earlier by Maury. Additionally, they estimated the effective temperatures of the components to be 17,000~K and 13,000~K.

Around the same time the light variations of $\mu^{1}~{\mbox{Sco}}$ were also measured by \citet{vanGent39} using photographic techniques. His 86~$m_{\mbox{\textit{\tiny pg}}}$ measurements showed variations for the primary and secondary eclipses of 0.30 and 0.19~magnitudes respectively -- similar to that obtained by Rudnick and Elvey and consistent with the original estimate made by Maury. There was, however, a systematic difference in the magnitudes made on the two different systems with van Gent's being 0.02~magnitudes brighter than those of Rudnick and Elvey (although both sets of measurements used $\mu^{2}~{\mbox{Sco}}$ as the comparison star). This difference was attributed to the significantly different effective wavelengths at which the two sets of measurements were made as there is only a very small difference between the colour indices of the variable and comparison star.

\begin{figure*}
\includegraphics[width=\textwidth]{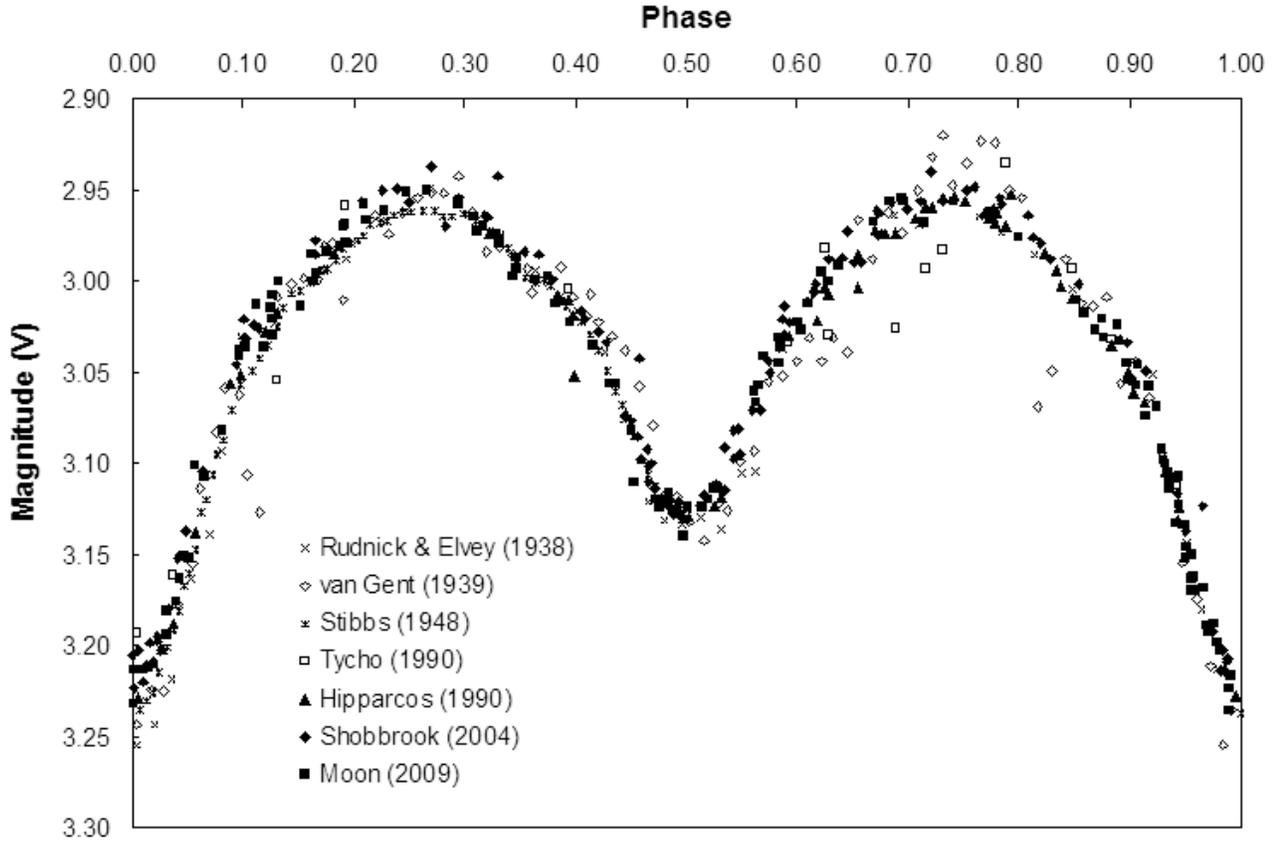}
\caption[Measured magnitudes of $\mu^{1}~{\mbox{Sco}}$.]{Measured magnitudes of $\mu^{1}~{\mbox{Sco}}$ as function of phase using an epoch of HJD~2449534.1770 and a period of 1.4462700~day. The typical error for a photoelectric measurement is no more than $\pm 0.01$ magnitude. For photographic measurements (i.e.\ those of van Gent) the typical error is $\pm 0.05$ magnitude.}
\label{fig:all_data}
\end{figure*}

For the magnitude versus phase plot of van~Gent there also appears to be a small shift in phase when it is compared to all the other photometric datasets. This shift corresponds to approximately~0.02 of the phase for the secondary minimum which equates to 40~minutes. It is not clear whether this is a real feature or a measurement error. When comparing the various photometric datasets (see Figure~\ref{fig:all_data}) his is the only dataset displaying this discrepancy.  An adjustment was thus made to all van~Gent measurements along similar lines to that discussed in \citet{MoonvanAntwerpen09}.

In his analysis \citet{vanGent39} noted the primary and secondary minima appeared to be symmetrically and equally spaced. By calculating the difference between the observed secondary minimum and the halfway point between consecutive primary minima, he showed the values for the eccentricity and longitude of periastron given by Maury were inconsistent with the observed light curve. Consequently he suggested the eccentricity is close to zero and considered a circular orbit for his subsequent analysis. This was subsequently confirmed through new spectroscopic observations (see \citealt{Stibbs48}). Using the observed light curve, along with a radii ratio adopted from Maury's estimate of line intensity ratios, van Gent determined the mass of the system to be 25.5~M$_{\sun}$ and the effective temperatures of the components to be 20,000~K and 12,500~K.

\citet{Stibbs48} results were derived from 358~photoelectric observations made with a gas-filled sodium cell operating in a broad spectral passband from 325~nm to 570~nm with an effective wavelength of 397~nm. Using these data, along with the ratio of luminosities of the two components determined by \citet{StruveElvey42}, he calculated orbital elements for the $\mu^{1}~{\mbox{Sco}}$ binary system and compared his results with those of \citet{RudnickElvey38} and \citet{vanGent39} showing that there was general agreement between them. Stibbs determined the mass of the system to be 23.2~M$_{\sun}$; $\mu^{1}~{\mbox{Sco}}$ was thus considered to comprise a primary component around 13 to 14~solar masses with a secondary around 9~solar masses in a relative orbit with a radius of a little over 10~million kilometres. The secondary was considered to be filling its Roche lobe and larger than the primary by about 10\%. The data tabulated were determined by assuming the measurements made were symmetrical about the \emph{epoch of the minimum under consideration, and then combined into normals containing ten observations, the normals being formed progressively at intervals of five observations.} This is in effect making the assumption that the orbit is circular. Stibbs then reduced his 358~observations to 72~values covering half a period.  Only the 72~measurements are available from the literature.

\citet{CesterFedel77} analysed data for 12~systems classified as being semi-detached binaries that presented what was, at that time, regarded as an evolutionary paradox because the less massive secondary component appeared to be in a more advanced stage of evolution than the primary. For their analysis of $\mu^{1}~{\mbox{Sco}}$ they used \citet{Stibbs48} data and determined masses and radii consistent with earlier works. In their discussion of this star \citet{CesterFedel77} remarked that the radius of the secondary, which appeared to be overflowing its Roche lobe, coupled with the high mass ratio was unusual for sd-binaries. Additionally, they regarded its position in HR diagram to be unusual as both components appeared to be lying on the main sequence. Noting that $\mu^{1}~{\mbox{Sco}}$ was a member of the Scorpius-Centaurus cluster whose age is at most $10^7$~years, \citet{CesterFedel77} suggested that this star may have just arrived on the main sequence.

\citet{SchneiderDarland79} also studied $\mu^{1}~{\mbox{Sco}}$ using Stibbs data and determined masses and radii generally consistent with earlier works. They evaluated the effective temperatures of the primary and secondary components to be 21,500~K and 16,200~K but, in contrast to earlier works, found the primary component to be about 5\% larger than the secondary.

\citet{Shobbrook04} undertook a program of fresh observations of those bright eclipsing binaries which had not been observed for many years with a view to establishing more recent epochs and more accurately determining their periods. Observations were made in the Str\"omgren $y$~band and transformed to $V$~magnitudes. Shobbrook determined a period of 1.446270(1)~day for $\mu^{1}~{\mbox{Sco}}$ and established a new epoch for its primary minimum of HJD~2449534.178 using his 112~new measurements, Hipparcos measurements and those taken by Stibbs 17,533~days earlier.

Hipparcos $H_p$ and Tycho $V_T$ magnitudes for $\mu^{1}~{\mbox{Sco}}$ were taken from the Hipparcos and Tycho catalogues \citep{ESA97} made available through the \vizier{} database \citep{OchsenbeinBauer00}. The Hipparcos data cover the period from 27~January~1990 to 22~February~1993, and the Tycho data from 10~February~1990 to 22~February~1993. In the analysis undertaken, only data where the value of the quality flag was less than or equal to two were used.

\citet{SticklandSahade96} noted the paucity of radial velocity data available for $\mu^{1}~{\mbox{Sco}}$ and undertook an analysis of spectroscopic data available from the IUE satellite \citep{IUE09}; in all 19~measurements for each component. They noted the signal of the secondary to be less than one third of that of the primary making measurements of radial velocity for both components difficult. Based on their analysis, they concluded that the orbit may have a small eccentricity of $0.019 \pm 0.017$. Additionally, there could be effects arising from tidal influences, gas streams or gravity darkening. \citet{SticklandSahade96} determined the system to be at an inclination of approximately 62$^{\circ}$, with a primary component of 8.6~M$_{\sun}$ and a secondary of 5.6~M$_{\sun}$, just filling its Roche lobe.

\citet{AriasSahade05} examined $\mu^{1}~{\mbox{Sco}}$ spectra for discrete UV features and, in the process, obtained radial velocity estimates from the 18~spectra they collected during July~1990. While stating their radial velocity estimates to be in very good agreement with those of \citet{SticklandSahade96}, they only provide a combined spectral plot the resolution of which is insufficient for a satisfactory estimate of velocity values to be made. Their results do, however, confirm those of \citet{SticklandSahade96}.

\begin{figure}
\includegraphics[width=1.0\columnwidth]{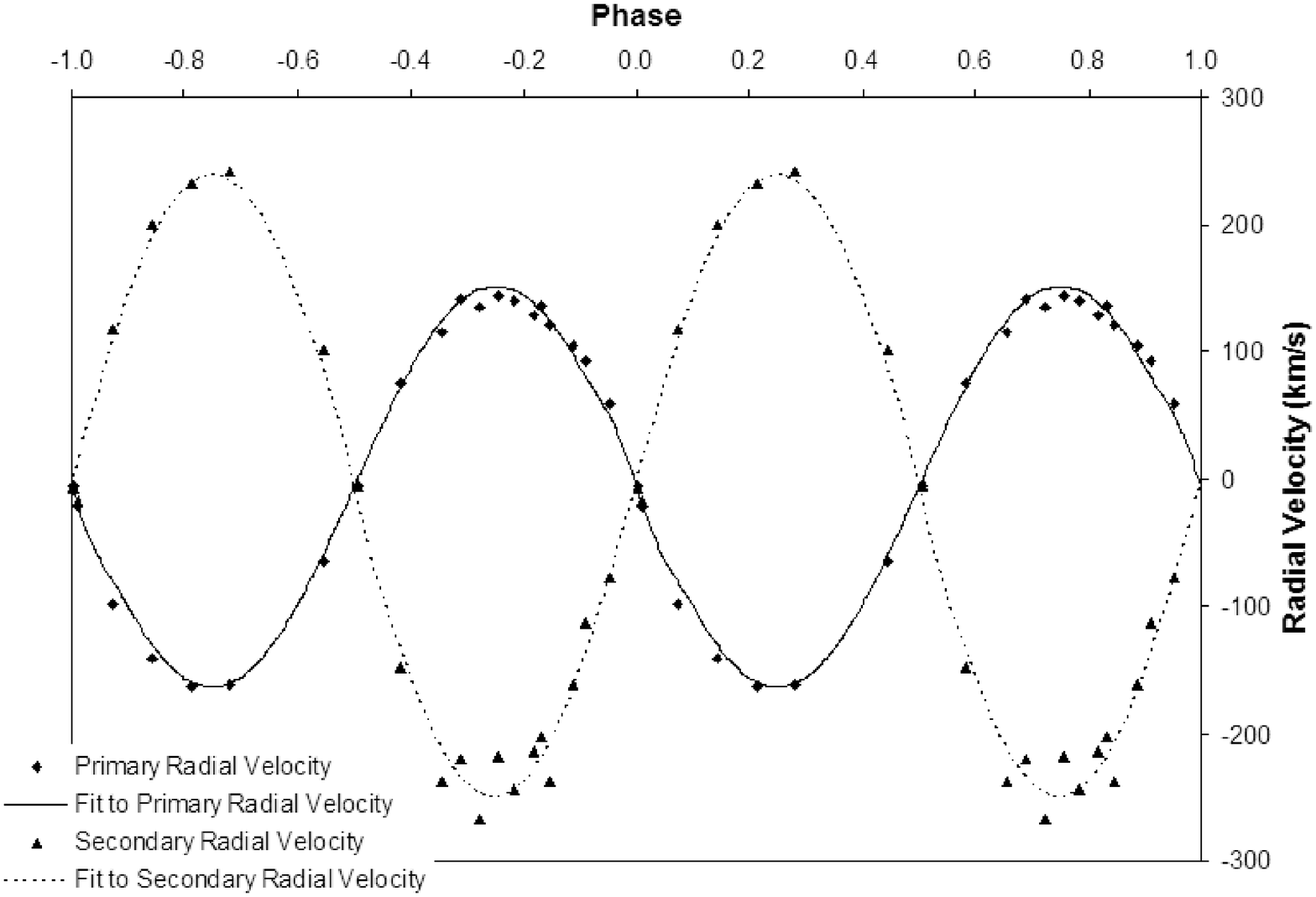}
\caption{Measured radial velocities of the components $\mu^{1}~{\mbox{Sco}}$ as function of phase using an epoch of HJD~2449534.1770 and a period of 1.446270~day.}
\label{fig:radial}
\end{figure}

Figure~\ref{fig:radial} displays the radial velocity data and resulting theoretical fit using \phoebe{}.

\section{Unpublished data}

\begin{table*}
\caption{The 133~new photometric measurements of Moon transformed to Johnson $V$ magnitudes and $B-V$ colour indices.}
\label{tab:newdata}
\begin{center}
\begin{tabular}{ccc|ccc|ccc|ccc}
\hline
{\bf HJD} & {\boldmath{$B-V$}} & {\boldmath{$V$}} & & {\bf HJD} & {\boldmath{$B-V$}} & {\boldmath{$V$}} & & {\bf HJD} & {\boldmath{$B-V$}} & {\boldmath{$V$}} \\
\hline
2453919.89054 & -0.190 & 3.056 &  & 2453993.95009 & -0.194 & 2.991 &  & 2454632.98615 &  & 3.126 \\
2453920.90439 & -0.185 & 3.000 &  & 2453999.93147 & -0.198 & 2.962 &  & 2454633.00490 &  & 3.124 \\
2453922.97861 & -0.186 & 3.057 &  & 2454305.89568 & -0.191 & 2.974 &  & 2454633.02295 &  & 3.123 \\
2453925.96041 & -0.192 & 3.000 &  & 2454321.98683 &        & 3.110 &  & 2454633.03893 &  & 3.114 \\
2453926.04097 & -0.188 & 2.956 &  & 2454322.91661 &        & 3.038 &  & 2454633.90836 &  & 3.021 \\
2453936.04875 & -0.195 & 3.027 &  & 2454322.94855 &        & 3.036 &  & 2454633.96114 &  & 3.000 \\
2453936.90008 & -0.191 & 2.969 &  & 2454322.96036 &        & 3.029 &  & 2454634.00489 &  & 2.979 \\
2453936.92578 & -0.184 & 2.958 &  & 2454325.90455 &        & 2.985 &  & 2454641.92347 &  & 2.967 \\
2453936.95147 & -0.201 & 2.961 &  & 2454325.92330 &        & 2.984 &  & 2454641.96166 &  & 2.954 \\
2453936.97994 & -0.190 & 2.951 &  & 2454325.94136 &        & 2.981 &  & 2454659.91927 &  & 3.013 \\
2453937.00772 & -0.191 & 2.950 &  & 2454326.90169 &        & 3.010 &  & 2454667.88825 & -0.176 & 2.995 \\
2453937.91113 & -0.198 & 3.033 &  & 2454326.92599 &        & 3.027 &  & 2454670.88805 & -0.211 & 2.956 \\
2453937.95071 & -0.184 & 3.058 &  & 2454326.93780 &        & 3.031 &  & 2454674.90930 &  & 3.124 \\
2453937.97154 & -0.204 & 3.100 &  & 2454326.95516 &        & 3.024 &  & 2454674.92180 & -0.192 & 3.116 \\
2453937.97571 & -0.177 & 3.109 &  & 2454326.96974 &        & 3.054 &  & 2454674.94264 & -0.205 & 3.131 \\
2453942.95107 & -0.187 & 2.997 &  & 2454326.99335 &        & 3.074 &  & 2454674.96486 & -0.190 & 3.124 \\
2453953.89400 & -0.176 & 3.108 &  & 2454327.01210 &        & 3.092 &  & 2454674.97319 & -0.204 & 3.120 \\
2453953.90650 & -0.189 & 3.134 &  & 2454327.03154 &        & 3.133 &  & 2454677.92436 & -0.204 & 3.060 \\
2453953.91483 & -0.184 & 3.150 &  & 2454327.04543 &        & 3.146 &  & 2454677.93686 & -0.193 & 3.041 \\
2453953.92942 & -0.171 & 3.168 &  & 2454329.93615 &        & 3.152 &  & 2454677.95908 & -0.215 & 3.036 \\
2453953.93775 & -0.184 & 3.192 &  & 2454329.94865 &        & 3.170 &  & 2454677.98130 & -0.207 & 3.022 \\
2453953.94678 & -0.183 & 3.198 &  & 2454331.89986 &        & 2.965 &  & 2454686.94657 & -0.218 & 2.976 \\
2453953.96344 & -0.195 & 3.223 &  & 2454331.95541 &        & 2.993 &  & 2454694.97020 & -0.202 & 2.987 \\
2453953.98080 & -0.209 & 3.232 &  & 2454331.98041 &        & 2.999 &  & 2454705.90601 &  & 3.046 \\
2453953.99816 & -0.194 & 3.212 &  & 2454332.00680 &        & 3.012 &  & 2454705.93031 &  & 3.069 \\
2453954.00649 & -0.206 & 3.210 &  & 2454332.02624 &        & 3.022 &  & 2454705.94351 &  & 3.105 \\
2453954.02316 & -0.193 & 3.194 &  & 2454334.90723 &        & 3.011 &  & 2454705.95809 &  & 3.122 \\
2453954.03913 & -0.203 & 3.163 &  & 2454334.94820 &        & 3.035 &  & 2454705.97614 &  & 3.162 \\
2453954.04538 & -0.196 & 3.151 &  & 2454344.92991 &        & 2.970 &  & 2454713.90109 &  & 3.056 \\
2453962.93005 & -0.186 & 2.970 &  & 2454344.96810 &        & 2.997 &  & 2454713.98928 &  & 3.140 \\
2453969.94401 & -0.195 & 3.176 &  & 2454346.89778 &        & 2.975 &  & 2454721.90033 &  & 3.168 \\
2453969.96068 & -0.196 & 3.152 &  & 2454346.94986 &        & 2.968 &  & 2454721.91977 &  & 3.203 \\
2453969.98081 & -0.185 & 3.107 &  & 2454355.90526 & -0.196 & 3.057 &  & 2454721.93505 &  & 3.216 \\
2453970.00373 & -0.198 & 3.082 &  & 2454355.93998 & -0.189 & 3.098 &  & 2454721.94963 &  & 3.213 \\
2453970.02664 & -0.195 & 3.041 &  & 2454355.96567 & -0.182 & 3.134 &  & 2454721.96213 &  & 3.213 \\
2453970.03567 & -0.198 & 3.036 &  & 2454357.91271 & -0.222 & 2.958 &  & 2454721.99199 &  & 3.181 \\
2453975.97748 & -0.202 & 2.966 &  & 2454357.93771 & -0.202 & 2.972 &  & 2454725.91176 &  & 2.956 \\
2453976.93712 & -0.194 & 3.021 &  & 2454357.96548 & -0.194 & 2.979 &  & 2454727.95323 &  & 3.014 \\
2453977.96271 & -0.197 & 3.045 &  & 2454374.92434 &        & 3.101 &  & 2454732.91666 &  & 3.031 \\
2453988.92904 & -0.206 & 2.996 &  & 2454377.91644 & -0.197 & 3.015 &  & 2454732.95346 &  & 3.012 \\
2453989.93102 & -0.191 & 3.017 &  & 2454381.92651 & -0.200 & 3.045 &  & 2454740.93121 &  & 3.008 \\
2453992.93491 & -0.202 & 3.114 &  & 2454609.95201 &        & 3.066 &  & 2454747.91673 &  & 3.163 \\
2453992.96338 & -0.202 & 3.170 &  & 2454632.93059 &        & 3.082 &  & 2454747.94520 &  & 3.188 \\
2453992.98490 & -0.182 & 3.189 &  & 2454632.96184 &        & 3.120 &  &  &  &  \\
2453993.01337 & -0.206 & 3.235 &  & 2454632.97434 &        & 3.122 &  &  &  & \\
\hline
\end{tabular}
\end{center}
\end{table*}

Photoelectric measurements of $\mu^{1}~{\mbox{Sco}}$ were made by one of the authors (Moon) from 3~July~2006 to 8~October~2008 using an Optec SSP-5A photometer attached to a permanently mounted 10-cm telescope that is housed in an observatory with a roll-off roof. For each measurement, the integration time is 10~seconds with each observation being the mean of 5~consecutive measurements. As the observatory is situated in an outer suburb of a major (Australian) city, the background sky was measured for each observation. When measuring through both $B$ and $V$ filters the sequence was $V_{\mbox{\sevensize star}}$, $B_{\mbox{\sevensize star}}$, $B_{\mbox{\sevensize sky}}$, $V_{\mbox{\sevensize sky}}$ \citep{OteroMoon06}. The resulting 133~new measurements of $V$ magnitude and 68~of $B-V$ were determined relative to $\mu^{2}~{\mbox{Sco}}$ (HR~6252; HD~151985; HIP~82545), the same comparison star used by previous observers. The check star, HR~6214 (HD~150742; HIP~81972), was the same as that used by Shobbrook. The magnitudes and colour indices obtained were transformed to the Johnson~$V$ band using well-determined transformation coefficients. These new measurements are presented in Table~\ref{tab:newdata}.

The photometric data on $\mu^{1}~{\mbox{Sco}}$ used in this analysis thus represent a comprehensive, and perhaps exhaustive, dataset comprising 512~measurements spanning 70~years.

\section{Analysis}

\subsection{Correction of photometric measurements to Johnson $V$ magnitudes.}

\begin{table*}
\caption{Particulars for the photometric bands of the various datasets used.}
\label{tab:corr}
\begin{tabular}{lccl}
\hline
Photometric Band & Effective Wavelength (nm) & FWHM (nm) & References \\
\hline
Yerkes ({\textit{KH cell}}) & ? ~ 460 & ? & \citet{Stebbins31} \\
Photovisual ($m_{\mbox{\textit{\tiny pv}}}$) & 560 & ? & \citet{vanGent39} \\
Johnson $V$ & 550 & 89 & \citet{Allen73} \\
Johnson $B$ & 440 & 98 & \citet{Allen73} \\
Stromlo ({\textit{Na cell}}) & 397 & ~ 120 & \citet{Stibbs48} \\
Str\"omgren $y$ & 547 & 23 & \citet{MermilliodHauck97} \\
Hipparcos $H_p$ & 480 & 230 & \citet{Bessell00} \\
Tycho $V_T$ & 510 & 105 & \citet{Bessell00} \\
\hline
\end{tabular}
\end{table*}

The $\mu^{1}~{\mbox{Sco}}$ photometric data assembled by the authors comprises measurements made with different photometric systems and is thus a heterogeneous dataset; the particulars of the various photometric systems are given in Table~\ref{tab:corr}. Corrections were applied to the individual datasets to adjust them to the Johnson~$V$ band. Owing to its line-of-sight alignment with $\mu^{1}~{\mbox{Sco}}$, and similarity in both magnitude and spectral type, $\mu^{2}~{\mbox{Sco}}$ is an ideal comparison star and has been used as such by all observers listed here with most $\mu^{1}~{\mbox{Sco}}$ measurements given simply as a magnitude difference relative to $\mu^{2}~{\mbox{Sco}}$. In measurements made by \citet{Shobbrook04}, and those by Moon, $\mu^{2}~{\mbox{Sco}}$ was used as the primary comparison star with HR~6214 used as a check star. The consistent use of $\mu^{2}~{\mbox{Sco}}$ as the comparison star makes the adjustment of the various datasets to the Johnson~$V$ band easier.

The GCPD~\citep{MermilliodHauck97} value of V=3.565 was chosen as the magnitude for $\mu^{2}~{\mbox{Sco}}$ to which the differences measured for $\mu^{1}~{\mbox{Sco}}$ were then applied. Colour corrections were not applied as there is only a small difference 0.002~magnitude in the $B-V$ colour indices of $\mu^{1}~{\mbox{Sco}}$ and $\mu^{2}~{\mbox{Sco}}$.

Transformations for $V_T$ and $H_P$ magnitudes to the $V$~band are given by \citet{Bessell00}. \citet{PlataisPourbaix03} also give a transformation of $V_T$ to $V$ as a function of $B-V$. Using these two sources it was calculated that any correction for the colour difference between $\mu^{1}~{\mbox{Sco}}$ and $\mu^{2}~{\mbox{Sco}}$ would be $<0.001$~magnitude. Zero-point corrections of 0.041 and 0.082 were thus applied to the $V_T$ and $H_P$ magnitudes respectively.

The various systems used will have different internal errors, however a typical value for the errors of transformed photoelectric measurements is $\pm0.01$~magnitude while errors for photographic photometry are typically $\pm0.05$~magnitude \citep{BuddingDemircan07}.

\subsection{Determining Effective Temperature from photometric indices or spectral types}

While separate spectra for the components of $\mu^{1}~{\mbox{Sco}}$ can be discerned, their spectral types remain somewhat uncertain. \citet{Maury20} first assigned a spectral type of B3 to both components but noted a difference in their line intensities which is discussed in detail by \citet{StruveElvey42}. The GCVS classifies the components of $\mu^{1}~{\mbox{Sco}}$ as B1.5V + B6.5V although \citet{SticklandSahade96} discuss the possibility of the secondary being of a type earlier than B6.

As the measured colour indices for $\mu^{1}~{\mbox{Sco}}$ are for the combined light of the two components, tables of effective temperature and absolute magnitude as a function of spectral type or colour index cannot be readily used. It is, however, useful to examine if suitable values for the temperatures of the two components can be estimated from the combined-light colour index so as to assist with the initial choice of these parameters for subsequent modelling.

Table~\ref{tab:mphoto} lists the GCPD \citep{MermilliodHauck97} mean photometric values in the $UBV$ and $uvby$ systems for $\mu^{1}~{\mbox{Sco}}$ (accessed online, 18 December 2008). Being at a distance of about 150~pc, interstellar reddening would be expected to be small for $\mu^{1}~{\mbox{Sco}}$. Using the $Q$ value calculated from the $UBV$ photometry, assuming a standard reddening law can be applied to the combined light of the components, and using the relationships given by \citet{BuddingDemircan07} gave an $E_{\mbox{\tiny\textit{B-V}}} = 0.02$. Additionally, from the $uvby$ data, an $E_{\mbox{\tiny\textit{b-y}}} = 0.014$ was calculated which is consistent with the calculated $E_{\mbox{\tiny\textit{B-V}}}$ value. A value of $(B-V)_{\mbox{\tiny\textit{0}}} = -0.232$ was thus adopted for the combined $B-V$ colour index of $\mu^{1}~{\mbox{Sco}}$.

With the data in Table~\ref{tab:colour} (assembled from various sources), the $B-V$ colour arising from combination of various spectral types was then explored using a spreadsheet. Within the stated error of the listed $B-V$, the modelling indicated that the primary could not be earlier than B1.5V or later than B2V. For a primary of B1.5V the colour index varied only slightly as the secondary was varied in type from B3V to B8V. Within the listed error for the colour index, the spectral type of the secondary thus remains indeterminate. Best estimates for the effective temperatures of the components of $\mu^{1}~{\mbox{Sco}}$ using the combined $B-V$ colour index, the spreadsheet modelling and the values given in Table~\ref{tab:colour} are thus 22,800~K for the primary and between about 12,000~K and 17,000~K for the secondary.

The error in $T_{\mbox{\tiny eff}}$ may be estimated from the errors in $\log(T_{\mbox{\tiny eff}})$ for the stars used by \citet{Flower96} to determine his $B-V$, $T_{\mbox{\tiny eff}}$ relationship. For the 26~stars listed with $-0.30 < B-V < -0.15$ a mean error in $\log(T_{\mbox{\tiny eff}})$ of 0.02 was calculated. This translates to an error of approximately $\pm1000$~K for an early B star.

Using the parallax of $6.51\pm0.91$~mas for $\mu^{1}~{\mbox{Sco}}$ given by \citet{vanLeeuwen07}, along with a $V\approx 3.0$ for the combined light of the components, an $M_{\mbox{\textit{\tiny V}}}\approx -2.9\pm0.3$ was calculated.  This is consistent with a spectral type for the primary of B1.5V to B2V and in the range B3V to B8V for the secondary.

\begin{table}
\caption{GCPD photometric values (mean) for $\mu^{1}~{\mbox{Sco}}$ in the $UBV$ and $uvby$ systems.}
\label{tab:mphoto}
\begin{center}
\begin{tabular}{rl}
$V$ = & 3.044 $\pm$ 0.023 \\
$B-V$ = & -0.212 $\pm$ 0.007 \\
$U-B$ = & -0.850 $\pm$ 0.005 \\
$b-y$ = & -0.089 \\
$m_1$ = & 0.078 \\
$c_1$ = & 0.103 \\
\end{tabular}
\end{center}
\end{table}

\begin{table}
\caption{Colours, absolute magnitudes and effective temperatures for main sequence B stars. $B-V$ indices are from \citet{Zombeck90}, $M_{\mbox{\tiny\textit{V}}}$ from \citet{Cox00}, $T_{\mbox{\tiny eff}}$ from \citet{Flower96}, and $b-y$ and $M_{\mbox{\tiny\textit{V}}}(b-y)$ are from \citet{Moon85}.}
\label{tab:colour}
\begin{center}
\begin{tabular}{cccccc}
\hline
Sp. Type & $B-V$ & $M_{\mbox{\tiny\textit{V}}}$ & $T_{\mbox{\tiny eff}}$ & $b-y$ & $M_{\mbox{\tiny\textit{V}}}(b-y)$ \\
\hline
B0 & -0.30 & -4.00 & 33620 & -0.120 & -3.75 \\
B1 & -0.26 &  & 24338 & -0.115 & -3.31 \\
B2 & -0.24 & -2.45 & 21261 & -0.092 & -2.08 \\
B3 & -0.20 &  & 16958 & -0.080 & -1.33 \\
B5 & -0.16 & -1.20 & 14203 & -0.068 & -0.97 \\
B6 & -0.14 &  & 13197 & -0.059 & -0.74 \\
B7 & -0.12 &  & 12368 &  &  \\
B8 & -0.09 & -0.25 & 11376 & -0.045 & -0.32 \\
B9 & -0.06 &  & 10612 & -0.037 & 0.22 \\
\hline
\end{tabular}
\end{center}
\end{table}

\subsection{Choice of Epoch}

The earliest published photoelectric measurements with an accompanying epoch are those by \citet{RudnickElvey38} where they applied a correction of -0.006~day to Maury's published epoch. \citet{vanGent39}, \citet{Stibbs48}, \citet{SticklandSahade96}, \citet{Danilekiewicz02} and \citet{Shobbrook04} also supply epochs based on the primary minimum. Table~\ref{tab:epochs} summarizes the various epochs listed. Shobbrook's epoch was selected as a starting point as not only was it the most recent, but a small adjustment in the order of -0.001~day (within Shobbrook's stated error of $\pm$0.002~day) resulted in a good alignment of the available datasets. By trying different values around that given by Shobbrook, an epoch for the primary minimum of HJD~2449534.17700(9) was selected as the value where there was the best alignment of the available photometric datasets. This was confirmed using the fitting routine within the \phoebe{} software.

\begin{table*}
\caption{Published epochs of primary minimum and periods. \citet{AriasSahade05} values based on \citet{Danilekiewicz02}.}
\label{tab:epochs}
\begin{center}
\begin{tabular}{lll}
\hline
Reference & Epoch of Minimum (HJD) & Period (days) \\
\hline
\citet{Maury20} & 2412374.434 & 1.44627 \\
\citet{RudnickElvey38} & 2428281.250 & 1.44627 \\
\citet{vanGent39} & 2428414.2978 & 1.4462683$\pm$0.0000004 \\
\citet{Stibbs48} & 2432001.0451 & 1.44627$\pm$0.000001 \\
\citet{SticklandSahade96} & 2448102.521 & 1.446271$\pm$0.0000044 \\
\citet{Shobbrook04} & 2449534.178 & 1.446270$\pm$0.000001 \\
\citet{AriasSahade05}  & 2432001.0475 & 1.44626876 \\
Proposed in this paper & 2449534.17700 & 1.4462700$\pm$0.0000005 \\
\hline
\end{tabular}
\end{center}
\end{table*}

\subsection{Determination of the Period}

The period listed in the GCVS is based on that given by \citet{Stibbs48}. More recent values are, however, available from \citet{Shobbrook04}, or \citet{AriasSahade05}. Table~\ref{tab:epochs} details the various periods given in the literature with quoted errors where available.

Starting with Shobbrook's period estimate, small adjustments were made until the datasets aligned. It was found that a period of 1.4462700(5)~days best represents the 512~measurements collected over the period of the available measurements which is about 70~years. Figure~\ref{fig:all_data} shows that there is excellent alignment between the various datasets using this refined value of the period along with the chosen epoch. The fitting routine within \phoebe{} confirmed this result.

\subsection{The System's Orbital Characteristics}

\phoebe{} was developed by Pr\v sa and Zwitter \citep{Prsa03,PrsaZwitter05c,PrsaGuinan08,Phoebe09} and, while it is based on the Wilson-Devinney model \citep{WilsonDevinney71}, it includes many recent theoretical developments \citep{KallrathMilone99}. Version 0.31a of the model, (c.\ 2008), was used in this analysis. This software package requires separate radial velocity data for each component; hence the extensive data of \citet{Maury20}, which gives a radial velocity for the components combined, was not used.

Using estimated spectral types for the components of $\mu^{1}~{\mbox{Sco}}$, some parameters can be fixed in advance of modelling of the system's orbital characteristics. Owing to the radiative nature of B-type stars, the albedos of both the primary and secondary components were set to one as were the values for gravity darkening. The synchronicity parameter for each star was also set to one. The option of an atmospheric model for the stars was enabled as was the option of a reflection effect (set to four reflections). Factors such as third light, opacity, extinction and starspots were not considered. The limb darkening model adopted is based on a logarithmic law and van Hamme's tables \citep{vanHamme93}.

$\mu^{1}~{\mbox{Sco}}$ is a member of the Scorpius-Centaurus OB Association \citep{PreibischMamajek08}, the nearest such association to the Sun. \citet{DOraziRandich09} have analyzed the metallicity distribution of open clusters within 500~pc of the Sun and note that the majority of the clusters with close-to-solar metallicity are part of the Local Association which includes the Scorpius-Centaurus association. The metallicity for the $\mu^{1}~{\mbox{Sco}}$ components was thus set to the solar value.

The type of eclipsing binary to be modelled by \phoebe{} is selected from a range of options. Initial attempts to fit the photometric data using some of the various options for the system failed to adequately represent features present within the data. It was found that the semi-detached option, with the secondary filling its Roche Lobe appeared to provide a solution that best represents all the photometric data. The analysis thus concentrated on this option although a detached option was also investigated as it was able to represent many of the key features as the size of the secondary approached its Roche limit.

\subsubsection{Mass Ratio, Masses, Semi-Major Axis and Inclination}

\begin{figure}
\includegraphics[width=1.0\columnwidth]{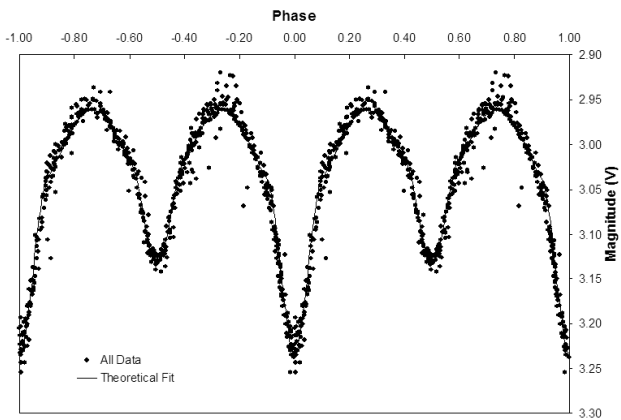}
\caption{\phoebe{}'s theoretical fit to all of the measured magnitudes of $\mu^{1}~{\mbox{Sco}}$ as function of phase using an epoch of HJD~2449534.1770 and a period of 1.4462700~day.}
\label{fig:fitdata}
\end{figure}

Using the radial velocity data of \citet{SticklandSahade96}, based on IUE data \citep{IUE09}, along with the \phoebe{} software, radial velocity curves were obtained and the mass ratio, semi-major axis and inclination determined. These values (slightly different to those found by \citet{SticklandSahade96}), are given in Table~\ref{tab:prop}. Figure~\ref{fig:fitdata} displays the radial velocity data from \citet{SticklandSahade96} and resulting theoretical fit.

\begin{table*}
\caption{$\mu^{1}~{\mbox{Sco}}$ system properties determined from modeling and analysis.}
\label{tab:prop}
\begin{center}
\begin{tabular}{ll}
\hline
\emph{Property} & \emph{Semi-Detached Model} \\
\hline
Primary Star's Spectral Type & B1.5V [22,800 $\pm$ 1000 K] \\
Primary Star's Mass & 8.49 $\pm$ 0.05 M$_{\sun}$ \\
Primary Star's Radius & 4.07 $\pm$ 0.05 R$_{\sun}$ \\
Primary Star's $T_{\mbox{\tiny eff}}$ (\phoebe)  & 23,725 $\pm$ 500 K \\
Primary Star's Potential & 3.85 $\pm$ 0.01 \\
\hline
Secondary Star's Spectral Type & B8 - B3 [12,000-17,000 K] \\
Secondary Star's Mass & 5.33 $\pm$ 0.05 M$_{\sun}$ \\
Secondary Star's Radius & 4.38 $\pm$ 0.05 R$_{\sun}$ \\
Secondary Star's $T_{\mbox{\tiny eff}}$ (\phoebe) & 16,850 $\pm$ 500 K \\
Secondary Star's Potential & 3.07 $\pm$ 0.01 \\
\hline
Orbital Inclination & 65.4 $\pm 1^{\circ}$  \\
Semi-Major Axis & 12.90 $\pm$ 0.04 R$_{\sun}$ \\
Eccentricity & 0.0 \\
Orbital Period & 1.4462700(5) day \\
Centre of Mass Velocity & -6.26 $\pm$ 0.04 km/s \\
Mass Ratio & 0.627 $\pm$ 0.004 \\
\hline
\end{tabular}
\end{center}
\end{table*}

While the values derived here are consistent with those of \citet{SticklandSahade96} to within the measurement errors, they are significantly different from those determined in earlier studies by Maury, van Gent, Rudnick and Elvey and Stibbs.

\subsubsection{Eccentricity, Argument of Periastron and Surface Potentials}

Using the available data noting the phase difference between the primary and the secondary minimums and the durations of the primary and secondary eclipses it is possible to determine the eccentricity. After normalizing the photometric datasets, a best-fit analysis using \phoebe{} derived an eccentricity of zero thus indicating the orbit is circular. Consequently the argument of periastron was set to zero along with the first time derivative of periastron. The circular orbit is consistent with previous analyses of $\mu^{1}~{\mbox{Sco}}$ and that tidal interactions in close binaries tend to circularize the orbit over time \citep{Hut81,KallrathMilone99}.

The surface potentials were estimated using the mass-ratio, radii and semi-major axis values established from fitting the radial velocity curves.

\subsubsection{System Properties}

The $\mu^{1}~{\mbox{Sco}}$ system was modelled iteratively starting with the recent data acquired by Moon (presented in this paper) and that of \citet{Shobbrook04} as they have the smallest photometric errors and together comprise half the the photometric measurements taken to date. Using initial estimates provided by this first phase, all available data was then used in the modelling of the $\mu^{1}~{\mbox{Sco}}$ binary system; the system properties thus determined are given in Table~\ref{tab:prop}. These values differ markedly from those determined by Maury, van Gent, Rudnick and Elvey and Stibbs. In particular the masses calculated are significantly less.

\begin{figure}
\includegraphics[width=1.0\columnwidth]{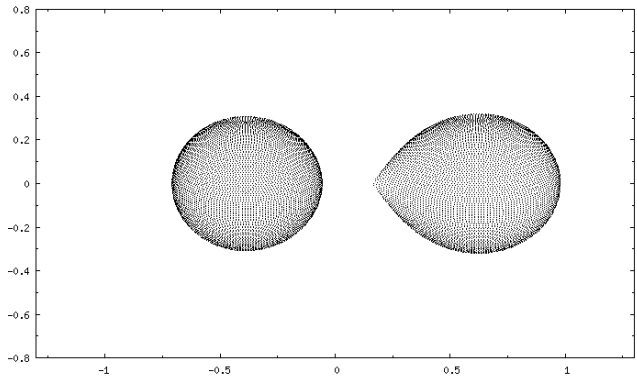}
\caption{A geometric configuration for $\mu^{1}~{\mbox{Sco}}$ at phase~0.25 based on the modelling and analysis undertaken, axes in units of semi-major axis.}
\label{fig:phase}
\end{figure}

Similarly, the effective temperatures determined by \phoebe{} were somewhat different to those determined in earlier studies but were found to be consistent with those estimated from the combined colour index. The theoretically-derived fit, using the parameters in Table~\ref{tab:prop} and all available photometric measurements, are plotted in Figure~\ref{fig:fitdata}. A geometric configuration for the stars at Phase~0.25 is presented in Figure~\ref{fig:phase}.

From the revised parallax for this star of $6.51\pm0.91$~mas \citep{vanLeeuwen07}, the semi-major axis of the orbit is calculated to be 0.39~mas. This star may thus be a suitable target for the Sydney University Stellar Interferometer~(SUSI).

\section{Conclusions}

The light variations of $\mu^{1}~{\mbox{Sco}}$ appear to be stable over the 70~years for which photometric data exists. Using the comprehensive dataset of more than 500~photoelectric measurements assembled here, the period was determined to be 1.4462700(5)~day. A suitable epoch based on the primary minimum is suggested to be HJD~2449534.17700(9).

Using all readily available data, and the \phoebe{} software package, the data for $\mu^{1}~{\mbox{Sco}}$ could be best represented by considering the system to be a semi-detached binary in which the size of the secondary is close to, or fills, its Roche Lobe. The system properties determined from comprehensive, iterative modelling using \phoebe{} indicate that the masses of the components may be somewhat less than previously thought. With the secondary appearing to fill its Roche lobe it maybe possible to obtain spectroscopic evidence of gaseous streams in the $\mu^{1}~{\mbox{Sco}}$ system noting that Doppler tomography has produced indirect images of gas flows in interacting binaries \citep{Richards06}. However, for Algol-type binaries the gaseous streams are faint relative to the main sequence primary star and their detection challenges current techniques. Improvements to techniques would likely be required to detect such gas flows in $\mu^{1}~{\mbox{Sco}}$.

There are currently only limited radial velocity data available for the $\mu^{1}~{\mbox{Sco}}$. Given that the radial velocity data provides initial values for parameters such as the semi-major axis, masses and radii, further data could be useful in improving the modelling of this binary system. Additionally, photometric measurements near phases 0.1, 0.4, 0.6 and 0.9 could aid refinement of the current theoretical model through better defining some apparent features.

With the semi-major axis of the orbit subtending 0.39~mas, this star may be a suitable target for the Sydney University Stellar Interferometer~(SUSI).

\section*{Acknowledgments}

Thank you to Dr A.C. Beresford for suggesting that this star was a worthwhile target for further study, Dr R.R. Shobbrook for discussions relating to his data and Associate Professor T. Bedding for helpful suggestions and his review of an earlier draft.

\hbadness=10000
\bibliography{bibliography}
\hbadness=1000

\end{document}